\documentclass[aps,prx,groupedaddress,twocolumn,showpacs,longbibliography,10pt]{revtex4-1}
\usepackage{amsmath}
\usepackage{amsthm, amssymb,bbold}    
\usepackage{braket}              
\usepackage{graphicx}
\usepackage{color}
\usepackage{times, verbatim}      
\usepackage{mathtools,tikz}
\usetikzlibrary{matrix}
\usepackage{natbib}
\usepackage{multirow}
\usepackage{hyperref}
\usepackage{subcaption}   
\newcommand{\eps}{\epsilon}

\definecolor{colour1}{rgb}{0.368417, 0.506779, 0.709798}
\definecolor{colour2}{rgb}{0.880722, 0.611041, 0.142051}
\definecolor{colour3}{rgb}{0,1,1}
\definecolor{colour4}{rgb}{0,1,0}
\definecolor{colour5}{rgb}{1,1,0}

\usepackage{amsthm}
\newtheorem*{theorem}{Theorem}
\hypersetup{
	colorlinks=true,
}

\begin{document}
	\newcommand{\tr}{\operatorname{tr}}
	\title{
		Homoclinic RG flows, or when relevant operators become irrelevant
	}
	
	\author{ Christian B. Jepsen$^{1}$, Fedor K. Popov$^{2}$  }
	\affiliation{$^{1}$ Simons Center for Geometry and Physics, Stony Brook University, Stony Brook, NY 11794}
	\affiliation{$^{2}$Department of Physics, Princeton University, Princeton, NJ 08544}

	\begin{abstract}
		We study an $\mathcal{N}=1$ supersymmetric quantum field theory with $O(M)\times O(N)$ symmetry.  Working in $3-\epsilon$ dimensions, we calculate the beta functions up to second loop order and analyze in detail the Renormalization Group (RG) flow and its fixed points. We allow $N$ and $M$ to assume general real values, which results in them functioning as bifurcation parameters. In studying the behaviour of the model in the space of $M$ and $N$, we demarcate the region where the RG flow is non-monotonic and determine curves along which %saddle-node and 
		Hopf bifurcations take place. At a number of points in the space of $M$ and $N$
		%at $N \approx 4.036,M \approx 2.945$ 
		we find that the model exhibits an interesting phenomenon: at these points the RG flow possesses a fixed point located at real values of the coupling constants $g_i$ but with a stability matrix $\left(\frac{\partial \beta_i}{\partial g_j}\right)$ that is not diagonalizable and has a Jordan block of size two with zero eigenvalue. %that has a pair of zero eigenvalues.
		Such points correspond to logarithmic CFTs and represent Bogdanov-Takens bifurcations, a type of bifurcation known to give rise to a nearby homoclinic orbit --- an RG flow that originates and terminates at the same fixed point. In the present example, we are able to employ analytic and numeric evidence to display the existence of the homoclinic RG flow.
		
	\end{abstract}
	
	\date{\today}
	
	
	\maketitle
	
	\section{Introduction}
	Since the classic review by Kogut and Wilson \cite{Wilson:1973jj} on the $\epsilon$ expansion and renormalization group (RG) flow, the general properties of RG flows have been the subject of active research. In the cases usually considered, once a theory starts flowing, it ends up at a fixed point where it is described by some conformal field theory (CFT). 
	From a general point of view, the equations describing instances of RG flow form systems of autonomous differential equations, and the properties of such systems and the kinds of flows they admit are well understood \cite{guckenheimer2013nonlinear,arnold2012geometrical,Gukov:2016tnp,feigenbaum1978quantitative}. In particular, dynamical systems can exhibit flows more peculiar than that between distinct fixed points, and Kogut and Wilson speculated in 1974 on the possibility of limit cycles as well as ergodic and turbulent behaviour in RG flow. 
	Since then, however, a number of monotonicity theorems have been proven that severely restrict the RG flow of unitary quantum field theories (QFTs). The first such theorem was Zamolodchikov's $c$-theorem \cite{zamolodchikov1986irreversibility}, which in two dimensions establishes a function that interpolates between central charges at CFTs and decreases monotonically along RG flow. Analogous theorems were proven in four dimensions ($a$-theorem) \cite{Komargodski:2011vj,Luty:2012ww} and  three dimensions  ($F$-theorem) \cite{Klebanov:2011gs,Jafferis:2011zi,Casini:2012ei}. The monotonicity implied by these theorems excludes the possibility of limit cycles, except for a loophole pointed out in \cite{morozov2003can,Curtright:2011qg}: multi-valued $c$ functions. This loophole had in fact been previously realized in certain deformed Wess-Zumino-Witten models \cite{Bernard:2001sc,LeClair:2003hj,Leclair:2003xj}, although these models required coupling constants to pass between infinity and minus infinity in order to realize cyclic RG flow. There are also examples of cyclics RG flow in quantum mechanics \cite{Glazek:1993qs,Glazek:2002hq,Bulycheva:2014twa,Gorsky:2013yba,LeClair:2002ux,Braaten:2004pg,Dawid:2017ahd}.
	
	Recently, ref. \cite{Jepsen:2020czw} put forward a QFT of interacting symmetric traceless matrices transforming under the action of the $O(N)$ group, while allowing $N$ to assume non-integer values. $O(N)$ models for non-integer $N$, an idea widely used in polymer physics \cite{de1979scaling}, had been previously given a formal definition in \cite{Binder:2019zqc}, which demonstrated the non-unitarity of these models.  Hence, the $c,a,F$-theorems are no longer valid and do not constrain the RG flow, and consequently  ref. \cite{Jepsen:2020czw} was able to show that the model studied therein possesses a closed limit cycle for $N$ slightly above $N_*\approx 4.475$. The main tool used to make this discovery was Hopf's theorem \cite{hopf1942bifurcation}, which guarantees the existence of a limit cycle in the vicinity of the codimension-one bifurcation known as the Andronov-Hopf bifurcation. 
	
	Turning to dynamical systems parameterized by two real numbers, codimension-two bifurcations can be used to prove the occurrence of yet other kinds of flow. Specifically, R.Bogdanov \cite{bogdanov} and F.Takens \cite{takens2001forced} have established powerful theorems by which, from properties of autonomous differential equations known only to second order in the dynamical variables, one can deduce the existence of homoclinic orbits, i.e. flow curves that connect a fixed point to itself. In addition to mild genericity conditions, the conditions that must be satisfied in order for the theorems to apply can be checked merely by studying the stability of fixed points, despite the fact that homoclinic orbits signal global bifurcations \cite{guckenheimer2013nonlinear} since they arise when a limit cycle collides with a saddle point.
	
	An interesting fact about homoclinic orbits is that they can be used to diagnose chaos. In applications of the theory of dynamical systems to physics, chaotic behavior \cite{cvitanovic2017universality} occurs in many instances, such as in turbulence \cite{ruelle1971nature,zakharov2012kolmogorov}, meteorology \cite{lorenz1963deterministic}  and even in scattering amplitudes in string theory \cite{Gross:2021gsj}. Usually, chaotic behaviour is proven via numerical investigations of concrete systems. One of the few analytical tools that can hint at the emergence of chaos is a theorem due to Shilnikov \cite{shilnikov1965case} that, for systems possessing homoclinic orbits, stipulates conditions by which to show they are chaotic. Therefore, one important step towards uncovering chaotic RG flow is to establish the existence of homoclinic RG flow.
	
	Brief previous mention of homoclinic RG flow can be found in \cite{Oliynyk:2005ak,Kuipers:2018lux}, which study non-linear sigma models and $QCD_4$ in the Veneziano limit. These references, however, mention the phenomenon solely for the purpose of pointing out its impossibility in those contexts. 
	
	In this short letter, we study a QFT with global $O(N)\times O(M)$ symmetry. Examining the RG flow of the theory as a function of $M$ and $N$, we determine the regime where the flow is non-monotonic. In this regime, we are able to establish the locations of a number of Bogdanov-Takens bifurcations, by which we are able to conclude that the theory exhibits homoclinic RG flow. In other words, the model contains fixed point with the peculiar property that a deformation by a {\it relevant} operator induces a flow that leads back to the original point: an RG flow where the IR and UV theories are one and the same. Homoclinic RG flow can be thought of as interpolating between the familiar type of RG flow (where a system flows from one fixed point to another) and the more exotic RG limit cycles (like limit cycles, homoclinic orbits are closed)%limiting cases of cycle where the periodicity is infinite)
	. In unitary QFTs, homoclinic RG flows are still forbidden by $c,a,F$-theorems, but a fixed point situated in a homoclinic orbit could possibly be described by a standard CFT, in contrast to fixed points that give rise to limit cycles by undergoing a Hopf bifurcation, and which require operators with complex scaling dimensions. 
	
	The method we adopt can be applied more broadly to find homoclinic orbits in two-parametric families of theories. We expect the phenomenon to be present in many other QFTs.

	\section{The model}
	\label{sec:model}
	
	We consider an $\mathcal{N}=1$ supersymmetric model of interacting scalar superfields $\Phi^i_{ab}$ that is invariant under the action of an $O(N)\times O(M)$ group in $d=3-\epsilon$ dimensions. The superfields are traceless-symmetric matrices with respect to the action of an $O(N)$ group and vectors under the action of an $O(M)$ group. There are four singlet marginal operators
	\begin{gather}
		O_1 = \tr\left[ \Phi^i \Phi^i \Phi^j \Phi^j\right], \quad O_2= \tr \left[ \Phi^i \Phi^j \Phi^i \Phi^j\right], \notag\\
		O_3 = \tr \left[\Phi^i\Phi^i\right]^2,\quad O_4 = \tr \left[\Phi^i \Phi^j\right] \tr \left[\Phi^i \Phi^j\right]\,, \label{eq:Ops}
	\end{gather}
	and so the full action is
	\begin{gather}
		S = \int d^d x d^2 \theta\Big( \tr \left[\Phi^i D^2_\alpha \Phi^i\right] + \sum_i g_i O_i \Big)\,.
		\label{action}
	\end{gather}
	% I COMMENTED OUT THE PARAGRAPH ABOUT $\mathcal{N}=2$, WHICH I DON'T THINK WE NEED.
	% Usually the supersymmetric theories has much more strict non-renormalizability theorems \cite{...}, that constrain further the RG flow. Thus for $\mathcal{N}=2$ the RG flow has the following form
	% \begin{gather}
	% \mu \frac{\partial g_i}{\partial \mu} = \left(-\epsilon + 4\gamma_\Phi\right) g_i, \label{eq:N2eq}
	% \end{gather}
	% where $\gamma_\Phi$ is a gamma function for the field $\Phi$. The form of the equations \eqref{eq:N2eq}
	% does not allow the RG limit cycles and other non-trivial topoligcal behaviours of the RG flow. Thus the eigenvalues of the stability matrix are either zero or $\lambda_\Phi = - \epsilon + 4 \gamma_\Phi + g_i \partial_i \gamma_\Phi$. Therefore continuing such theories to non-integer parameters we would not still have RG limit cycles. For instance, the eigenvalues But since for $\mathcal{N}=1$ we do not have such constraint we still can get the RG limit cycles.
	
	The RG flow of this model is gradient, meaning that there exists a function $F$ of the couplings and a four-by-four matrix $G_{ij}$ such that the beta functions of the theory satisfy the equation
	\begin{gather}
		\beta_i = \mu\frac{d g_i}{d \mu} = G_{ij} \frac{\partial F}{\partial g_j}\,.
	\end{gather}
	If $G_{ij}$ is positive or negative definite, this equation implies that $F$ changes monotonically with the RG flow, so that cyclic and homoclinic flow lines are impossible. By explicit computation to leading order in perturbation theory, we find that the metric has determinant
	\begin{gather}
		\det G  = \frac{1}{4}(M-1)^2(M+2)^2(N-3) \times\notag\\
		\times(N-2)^2(N+1)^2(N+4)^2(N+6)\,. \label{eq:det}
	\end{gather}
	We list the beta functions and the components of the metric in appendix \ref{sec:betas}. The zeroes in $\det G$ occur because of linear relations among the four operator of the theory at special values of $M$ and $N$, and their presence indicates that eigenvalues change sign as $N$ and $M$ are varied.
	Indeed one can check that the metric is sign-indefinite if $M\in\left(-2,1\right)$ or $N\in \left(-6,3\right)$, so that unusual RG flows are possible in this regime, and operators may develop complex scaling dimensions at real fixed points, which, in the terminology of \cite{Jepsen:2020czw}, are then termed "spooky". At integer values of $N$ and $M$, such operators are identically zero owing to the linear relations between the operators. The situation is closely analogous to the occurrence of evanescent operators at non-integer spacetime dimensions \cite{Collins:1984xc,Bos:1987fb,Dugan:1990df,Gracey:2008mf,Hogervorst:2014rta,Hogervorst:2015akt}.
	
\begin{figure}
		\begin{center}
			\includegraphics[width=0.45\textwidth]{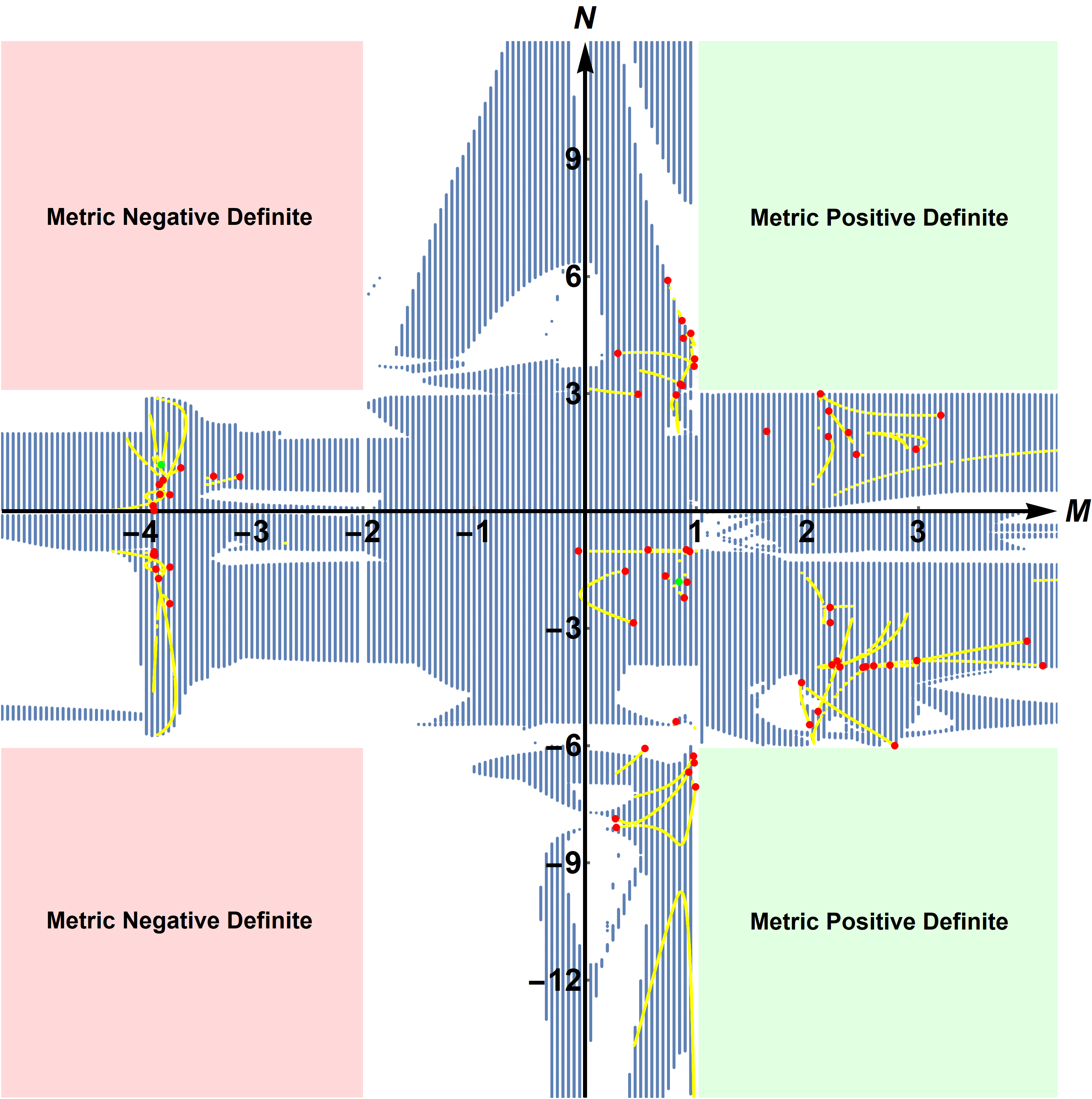}
		\end{center}
		\caption{ Bifurcation diagram in the space of $M$ and $N$. Spooky fixed points are present in the areas coloured in \textcolor{colour1}{blue}. Curves in \textcolor{colour5}{yellow} signify Andronov-Hopf bifurcations. Points coloured in \textcolor{red}{red} mark the locations of Bogdanov-Takens bifurcations, and points in \textcolor{green}{green} indicate Zero-Hopf bifurcations.}
		\label{fig:cross}
	\end{figure}

	In the following, we will allow $M$ and $N$ to assume general real values. In consequence of this analytic extension, we are able to observe Hopf bifurcations taking place in the model along various curves in the space of $M$ and $N$. See fig. \eqref{fig:cross}. But while Hopf bifurcations are a type of codimension-one bifurcation widely found in one-parameter systems of autonomous systems of differential equations, we are dealing with a two-parameter system, and such systems are capable of exhibiting a richer variety of flows. The possible codimension-two bifurcations can be classed into five types \cite{arnold2012geometrical,guckenheimer2013nonlinear} -- Bautin, Bogdanov-Takens, cusp, double-Hopf, and zero-Hopf -- which signal different kinds of flow not present in generic one-parameter systems. As we shall now see, some of these possibilities are realized by the QFT with action \eqref{action}.

	\section{Bogdanov-Takens Bifurcation}
	\label{sec:BT}
	A Bogdanov-Takens bifurcation occurs generically when, at a fixed point, two eigenvalues of the stability matrix $\left(\frac{\partial \beta_i}{\partial g_j}\right)$ tend to zero as two bifurcation parameters $M$ and $N$ are appropriately tuned. % To a pair of critical values $N_*,M_*$
	%In other words,
	The following equations must then be satisfied:
	\begin{gather}
		\beta_i(g_i,N,M) = 0\,, \quad \det \left(\frac{\partial \beta_i}{\partial g_j}\right) \left(g_i,N,M\right)=0\,, \label{eq:BTcon}\\
		\tr\left[\textstyle\bigwedge^3\left( \frac{\partial \beta_i}{\partial g_j}\right)\right] \equiv \det \left(\frac{\partial \beta_i}{\partial g_j}\right) \tr\left[ \left(\frac{\partial \beta_i}{\partial g_j}\right)^{-1} \right] = 0\,. \nonumber
		%\tr\left[\left( \frac{\partial \beta_i}{\partial g_j}\right)\right]^3 - 3  \tr\left[\left( \frac{\partial \beta_i}{\partial g_j}\right)\right]  \tr\left[\left( \frac{\partial \beta_i}{\partial g_j}\right)^2\right] + \notag\\ + 2  \tr\left[\left( \frac{\partial \beta_i}{\partial g_j}\right)^3\right]=0\,. \notag
	\end{gather}
	Written in the form \eqref{eq:BTcon}, we see that the conditions for a BT bifurcation are polynomial equations in $g_i$, $M$, and $N$, and so by B\'ezout's theorem there exist at most a finite number of points that satisfy these conditions. We refer to such points as Bogdanov-Takens (BT) points. For the QFT we are studying perturbatively, it can be verified that the beta functions exhibit several such points, as shown in figure \ref{fig:cross}. 
	Their existence can be checked to high numerical accuracy with the use of standard programs, e.g. PyDSTool \cite{pydstool}. 
	Higher-loop contributions will provide corrections to the precise locations of these points, but as long as we take $\epsilon$ to be sufficiently small,  higher-order corrections will not alter the number or qualitative behaviour of BT points. %One of the BT points is located at the integer values $M=2,\, N=3$, right on the boundary between monotic and non-monotonic RG flows.
	
	While two eigenvalues tend to zero as we approach a BT point, right at the BT point itself we do not typically have a pair of eigenvectors with zero eigenvalues, the reason being that in this same limit, the two respective eigenvectors usually become linearly dependent. Rather, the stability matrix at a BT point has a Jordan block of size two with zero eigenvalue (see \eqref{eq:numstabmatrix} in appendix \ref{sec:deriv}). This means that the theory at the BT point possesses two operators $\mathcal{O}_{1,2}$ such that the generator $D$ of dilatations acts in the following way
	\begin{gather}
		D \mathcal{O}_1 = d \mathcal{O}_1\,,\quad D\mathcal{O}_2 =d\mathcal{O}_2+\mathcal{O}_1\,.
	\end{gather}
	The possibility of indecomposable representations of the conformal group was extensively studied in \cite{Gurarie:1993xq,Hogervorst:2016itc}. The upshot is that the BT theory constitutes a logarithmic CFT containing {\it generalized} marginal operators $\mathcal{O}_{1,2}$. In consequence, BT theories are non-unitary and we have
	\begin{gather*}
		\braket{\mathcal{O}_2(0) \mathcal{O}_2(x)} =  -\frac{2 k_\mathcal{O} \log \left|x\right|}{\left|x\right|^{2d}}\,,\quad   \braket{\mathcal{O}_1(0) \mathcal{O}_2(x)} =  \frac{ k_\mathcal{O}}{\left|x\right|^{2d}}\,,
	\end{gather*}
	for some constant $k_\mathcal{O}$.
	\begin{figure}
		\begin{center}
			\includegraphics[width=0.4\textwidth]{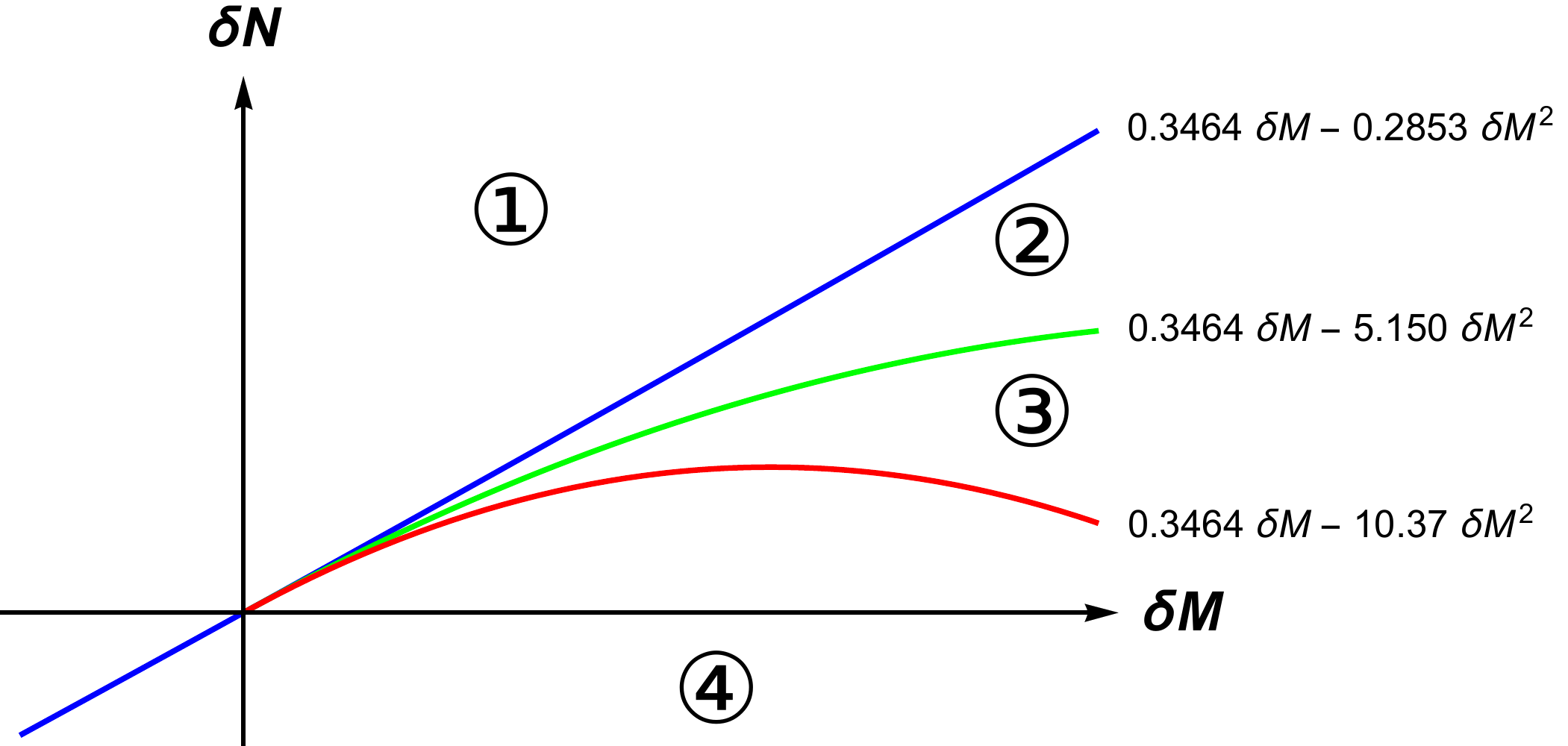}
		\end{center}
		\caption{Bifurcation diagram around the Bogdanov-Takens bifurcation at $(M,N)=(M^\ast,N^\ast)\approx (0.2945,4.036)$. $\delta M = M-M^\ast$, $\delta N=N-N^\ast$. The \textcolor{blue}{blue} curve represents a saddle node bifurcation, the \textcolor{green}{green} curve represents a Hopf bifurcation, and the \textcolor{red}{red} curve represents a saddle homoclinic bifurcation. At the origin, these three codimension-one bifurcations coalesce.}
		\label{fig:bifurcationDiagram}
	\end{figure}

	\begin{figure*}
		
		\begin{subfigure}{0.45\textwidth}
			\begin{center}
				\includegraphics[width=0.7\textwidth]{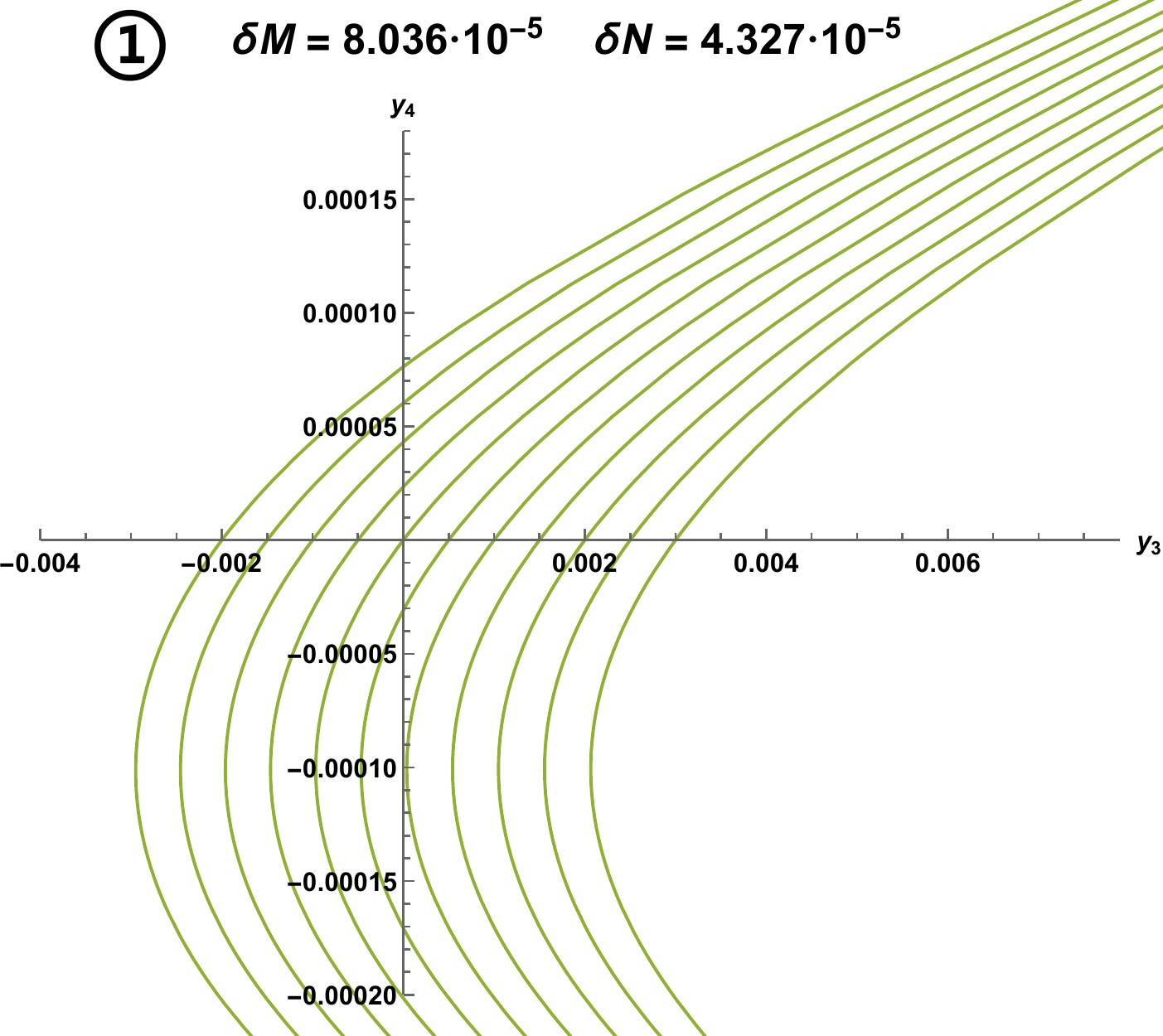}
			\end{center}
			\caption*{RG flow in region \raisebox{.5pt}{\textcircled{\raisebox{-.9pt} {1}}}. In this regime, there are no fixed points near the origin, and all flow curves swerve downwards in the IR.}
		\end{subfigure}
		\begin{subfigure}{0.45\textwidth}
			\begin{center}
				\includegraphics[width=0.7\textwidth]{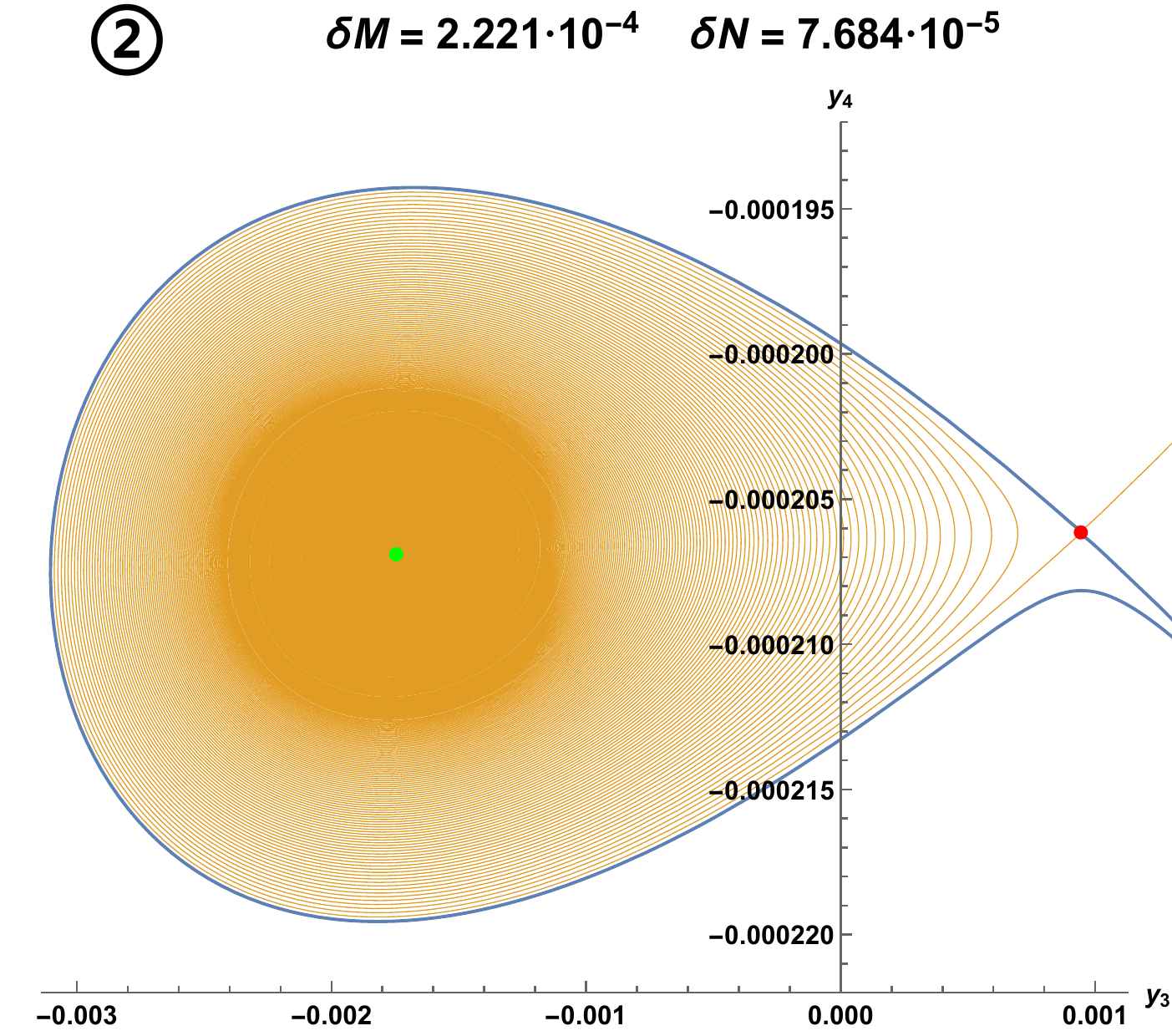}
			\end{center}
			\caption*{RG flow in region \raisebox{.5pt}{\textcircled{\raisebox{-.9pt} {2}}}. The fixed point marked in \textcolor{red}{red} is a saddle point. The \textcolor{colour1}{blue} curves flow outwards from this point in the IR, while the \textcolor{colour2}{orange} curves flow inward. The fixed point marked in \textcolor{colour4}{green} is IR unstable.  }
		\end{subfigure}
		\begin{subfigure}{0.45\textwidth}
			\begin{center}
				\includegraphics[width=0.7\textwidth]{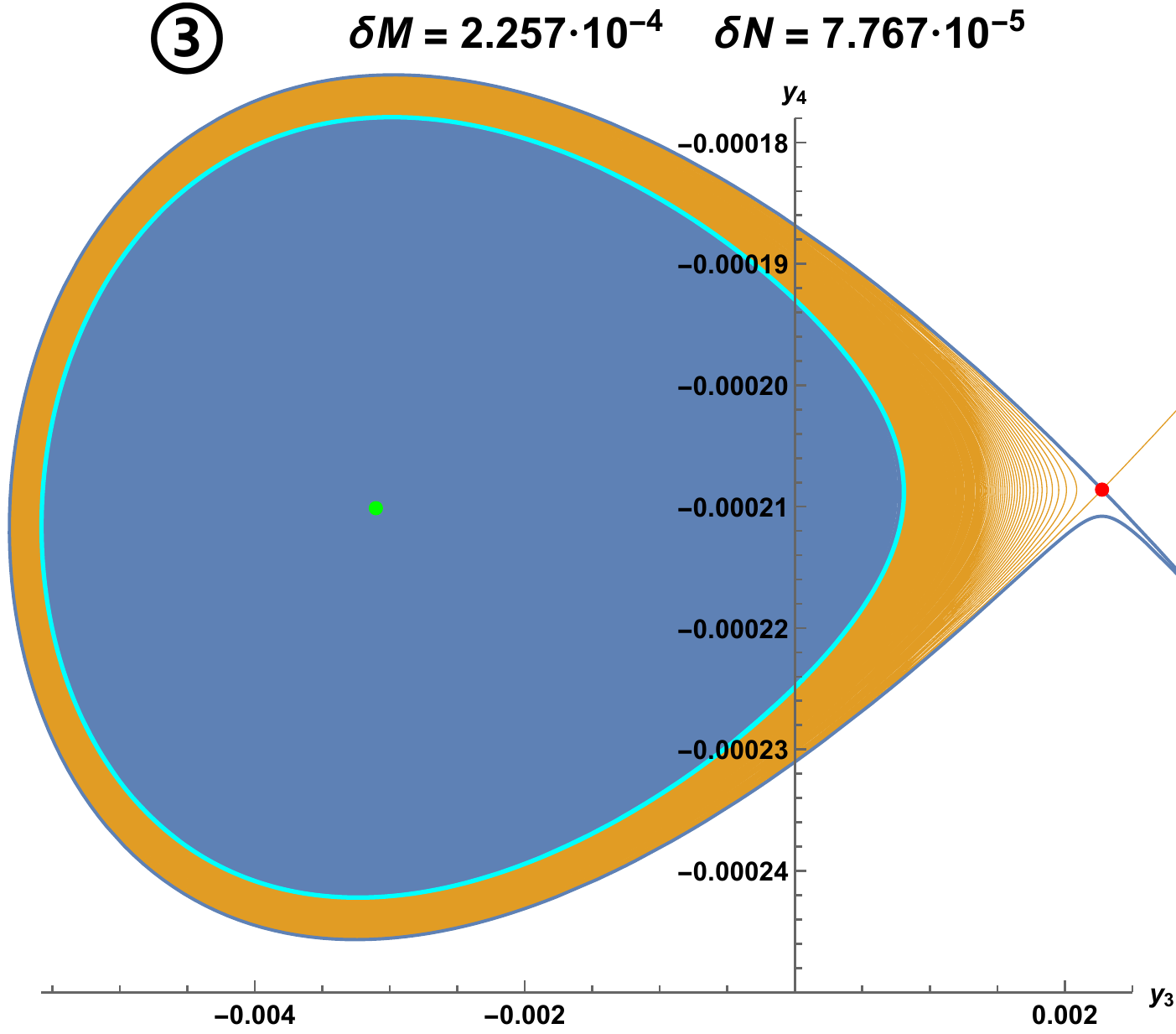}
			\end{center}
			\caption*{RG flow in region  \raisebox{.5pt}{\textcircled{\raisebox{-.9pt} {3}}}. In passing from region \raisebox{.5pt}{\textcircled{\raisebox{-.9pt} {2}}} to \raisebox{.5pt}{\textcircled{\raisebox{-.9pt} {3}}}, the fixed point marked in \textcolor{colour4}{green} has undergone a Hopf bifurcation and is now IR stable. An IR-repulsive limit cycle, marked in \textcolor{colour3}{cyan}, separates the two fixed points.}
		\end{subfigure}
		\begin{subfigure}{0.45\textwidth}
			\begin{center}
				\includegraphics[width=0.7\textwidth]{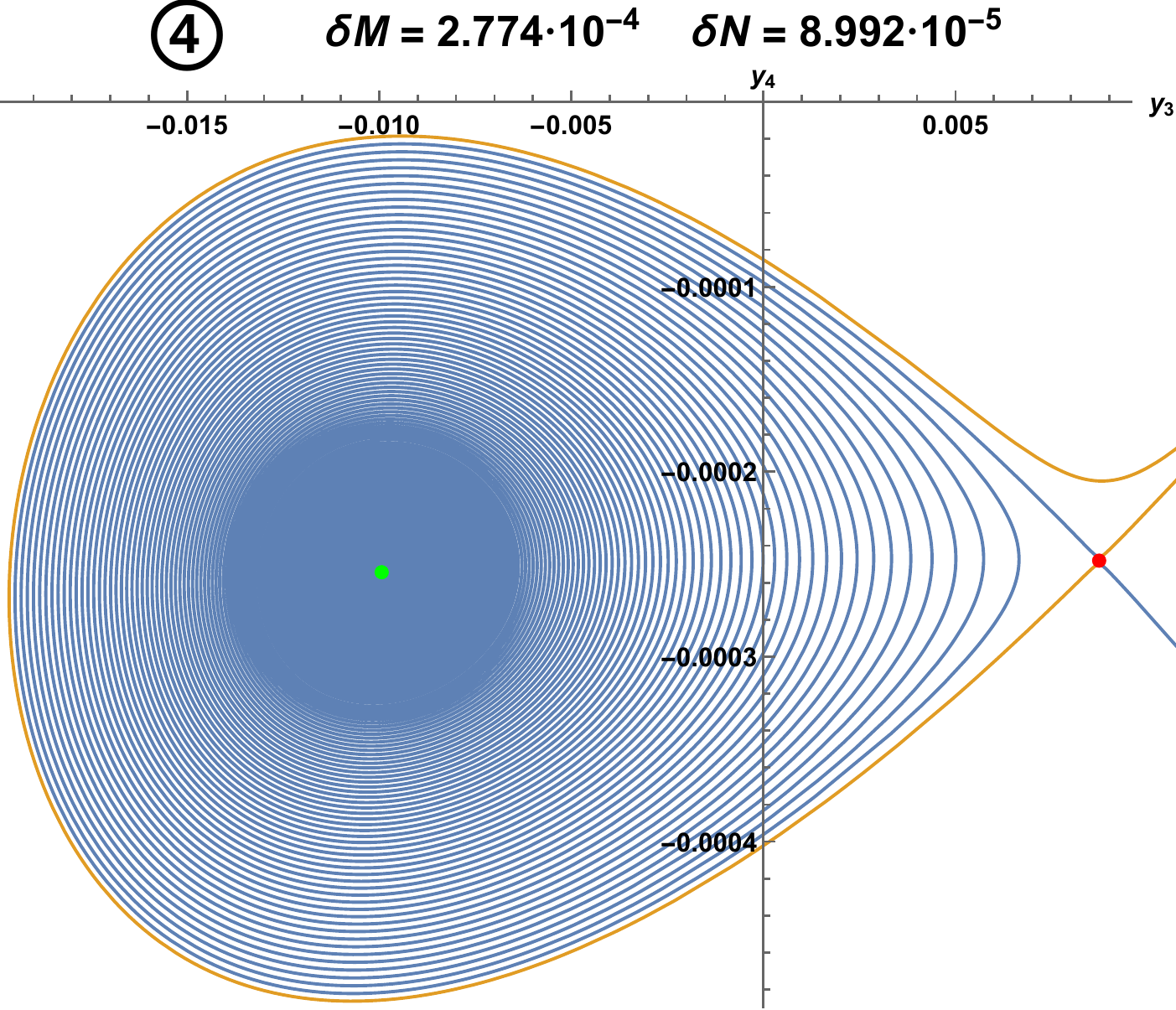}
			\end{center}
			\caption*{RG flow in region \raisebox{.5pt}{\textcircled{\raisebox{-.9pt} {4}}}. In passing from region \raisebox{.5pt}{\textcircled{\raisebox{-.9pt} {3}}} to \raisebox{.5pt}{\textcircled{\raisebox{-.9pt} {4}}} the limit cycle collided with the \textcolor{red}{red} fixed point in a homoclinic bifurcation. }
		\end{subfigure}
		\caption{The topologically distinct types of RG flow in the vicinity of the Bogdanov-Takens bifurcation at $M=M^\ast \approx 0.2945$ and $N=N^\ast \approx 4.036$. The variables $y_3$ and $y_4$ are linear combinations of the four coupling constants $g_i$, with precise definitions given in appendix \eqref{sec:deriv}, and $\delta M = M - M^\ast$, $\delta N = N - N^\ast$.}
		\label{fig:4regimes}
	\end{figure*}
	The conditions \eqref{eq:BTcon} are not entirely sufficient to guarantee a BT bifurcation. One must also require smoothness and a set of inequalities that are generically true. Violations of the inequalities typically require fine-tuning of additional parameters and signal bifurcations of codimension higher than two. Incidentally, at the integer values $M=2$ and $N=3$, right on the boundary of the regimes with monotonic and non-monotonic RG flows, we observe a fixed point that satisfies \eqref{eq:BTcon}, but which fails to meet these genericity requirements and for this reason is not described by a logarithmic CFT.
	
	In appendix \ref{sec:deriv} we give the precise statement of the Bogdanov-Takens bifurcation theorem, and we explicitly check that it applies to an example of a BT point in the QFT we are studying, situated at  $M \approx 0.2945$ and $N \approx 4.036$. What this means is that we can transform the beta functions near the BT point into a particularly simple form, known as Bogdanov normal form: 
	\begin{gather}
		\left\{\begin{matrix*}[l]
			\dot{\eta}_1=\eta_2\,, \\
			\dot{\eta}_2 = \delta_1+\delta_2 \eta_1+\eta_1^2 + s \eta_1 \eta_2 + \mathcal{O}\left(\left|\eta \right|^3\right)\,,\\
			\dot{\eta}_i=\lambda_i \eta_i\quad \text{for }i>2\,,
		\end{matrix*}\right. \label{eq:BTsys}
	\end{gather}
	where $s=- 1$, and $\delta_{1,2}$ are functions of $N$ and $M$ that vanish right at the BT point. 
	
	By bringing the system into normal form, we can use the equations \eqref{eq:BTsys} to determine the behaviour of the system for small enough $\delta_1$ and $\delta_2$. In particular, we can constrain ourselves to studying the surface where only $\eta_1$ and $\eta_2$ are non-zero, noting that the dynamics in the transverse directions $\eta_3$ and $\eta_4$ are quite simple. Depending on the values of $\delta_1$ and $\delta_2$, the flow of $\eta_{1,2}$ falls into different topological types. The classification can be found in textbook \cite{kuznetsov2013elements} and amounts to the following. In the vicinity of the BT point at $\delta_1=\delta_2=0$, there are four regimes with qualitatively different flows:
	
	-- Regime \raisebox{.5pt}{\textcircled{\raisebox{-.9pt} {1}}}: The flow has no fixed point.
	
	In the other three regimes, the flow has two fixed-points, which we will label left and right. The right point is always a saddle point.
	
	-- Region \raisebox{.5pt}{\textcircled{\raisebox{-.9pt} {2}}}: The left point is unstable, and all flow lines starting near it terminate at the right fixed point.
	
	-- Region \raisebox{.5pt}{\textcircled{\raisebox{-.9pt} {3}}}: The left point is now stable, and a repulsive limit cycle separates the two fixed points.
	
	-- Region \raisebox{.5pt}{\textcircled{\raisebox{-.9pt} {4}}}: The left point is still stable, but the limit cycle has disappeared. Some flow lines starting near the right fixed point terminate at the left fixed point.
	
	In the case of the BT point at $(M,N) \approx (0.2945,4.036)$, the locations of these four adjoining regimes, as computed in appendix \eqref{eq:BTsys}, is shown in figure \ref{fig:bifurcationDiagram}. And the RG flow in each regime is depicted in figure \ref{fig:4regimes}.
	
	The four regimes are separated by different codimension-one bifurcations. Region \raisebox{.5pt}{\textcircled{\raisebox{-.9pt} {1}}} is demarcated from regions \raisebox{.5pt}{\textcircled{\raisebox{-.9pt} {2}}} and \raisebox{.5pt}{\textcircled{\raisebox{-.9pt} {4}}} by a saddle-node bifurcation happening at $\delta_1=\frac{1}{4}\delta_2^2$. Regions \raisebox{.5pt}{\textcircled{\raisebox{-.9pt} {2}}} and \raisebox{.5pt}{\textcircled{\raisebox{-.9pt} {3}}} are separated by an Andronov-Hopf bifurcation along the the half-curve $\delta_1=0$, $\delta_2< 0$. And regions \raisebox{.5pt}{\textcircled{\raisebox{-.9pt} {3}}} and \raisebox{.5pt}{\textcircled{\raisebox{-.9pt} {4}}} are separated by a saddle homoclinic bifurcation along $\delta_1=-\frac{6}{25}\delta_2^2+\ldots, \, \delta_2 <0$.
	
	\begin{figure}
		\centering
		\includegraphics[scale=0.6]{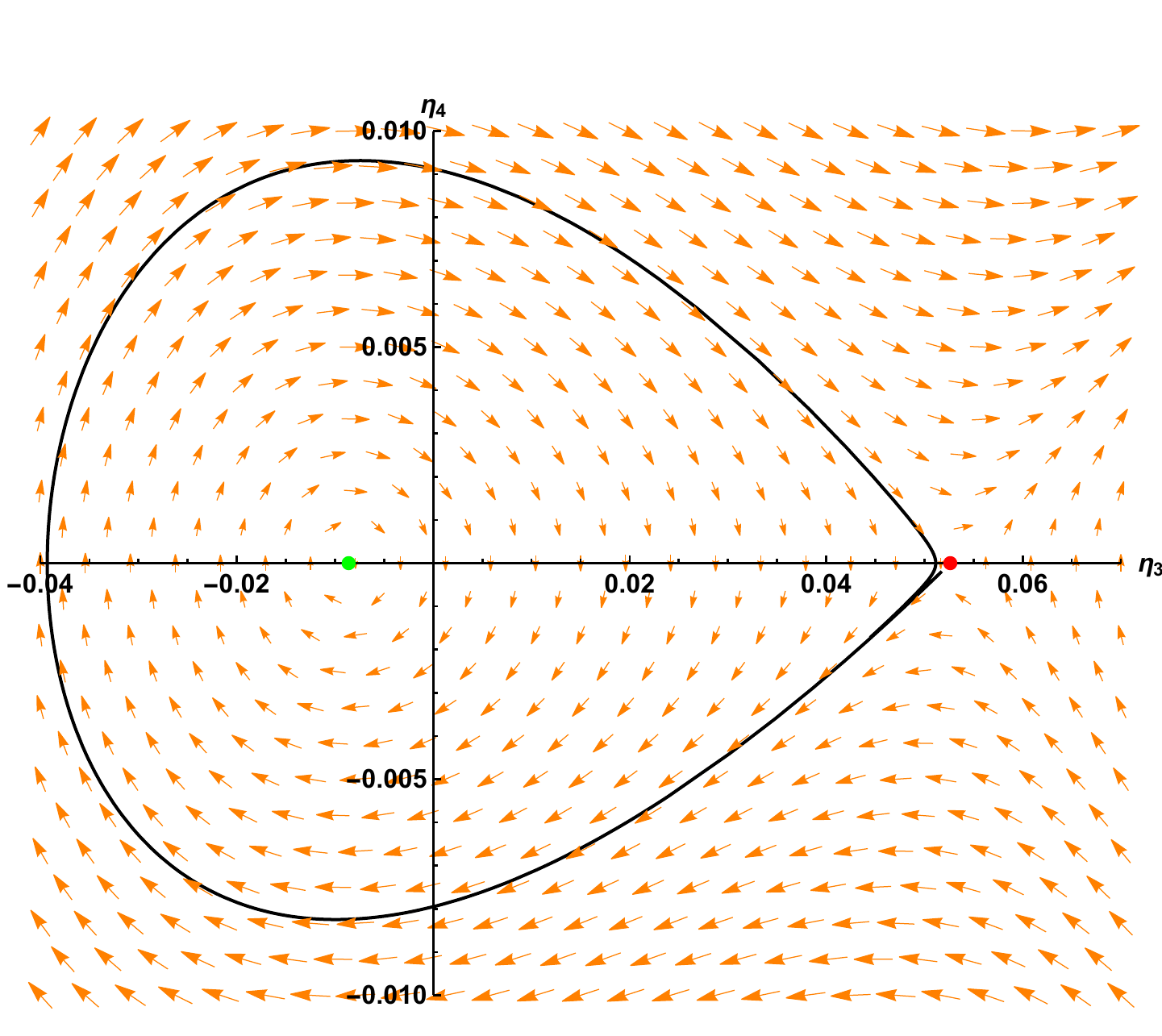}
		\caption{Flow diagram for a dynamical system containing a homoclinic orbit (marked in \textbf{black}), ie. a flow line that starts and ends at the same point. The system is described by equations \eqref{eq:A10} with parameters $\delta_1=-0.000453178$ and $\delta_2=-0.0440214$. The \textcolor{red}{red} and \textcolor{green}{green} dots indicate fixed points. The \textcolor{green}{green} dot is a "spooky" fixed point. The theory at the \textcolor{red}{red} dot is a homoclinic CFT.}
		\label{fig:homoclorbit}
	\end{figure}
	
	A saddle-node bifurcation corresponds to the collision and disappearance of two equilibria in dynamical systems. The phenomenon has been observed in a number of cases of RG flow, it happens for instance in in the critical $O(N)$ model \cite{fei2015three} and in prismatic models \cite{Giombi:2018qgp}, and has been proposed to occur in $QCD_4$ \cite{Gies:2005as,kaplan2009conformality,Gorbenko:2018ncu,Kuipers:2018lux}.
	
	An Andronov-Hopf bifurcation represents a change of stability at a fixed point that has complex eigenvalues. The flow near the fixed point changes between spiraling inwards and spiraling outwards and gives birth to a limit cycle. In the context of RG, this bifurcation was recently studied in \cite{Jepsen:2020czw}. 
	
	The most interesting and new phenomenon associated to the model of the present paper happens along the homoclinic bifurcation line. Here the flow exhibits what is known as a homoclinic orbit.

	\section{Homoclinic RG flow}
	
	A homoclinic orbit is a flow line that connects a stable and an unstable direction of a saddle point. Figure \eqref{fig:homoclorbit} depicts the kind of homoclinic orbit generated by a BT bifurcation, with the saddle point marked by a \textcolor{red}{red} dot. The homoclinic orbit is seen to envelop another fixed point marked in \textcolor{green}{green}. In a QFT context, the \textcolor{green}{green} point is "spooky": the couplings are real, but the eigenvalues of the stability matrix $\left(\frac{\partial \beta_i}{\partial g_j}\right)$ have non-zero imaginary parts. In contrast to such spooky points, and to complex CFTs \cite{Gorbenko:2018ncu,Gorbenko:2018dtm}, the 
	\textcolor{red}{red} saddle point is associated to real couplings and real eigenvalues of the stability matrix. These eigenvalues are small and have opposite signs: $\lambda_1,-\lambda_2 \ll 1$. The positive eigenvalue corresponds to a slightly relevant operator $\mathcal{O}_1$ with dimension $\Delta_1=d+\lambda_1 > d$, and the negative eigenvalue to a slightly irrelevant operator $\mathcal{O}_2$ with dimension $\Delta_2=d + \lambda_2 < d$. In this sense, the \textcolor{red}{red} saddle point corresponds to a real CFT.
	
	Standard RG lore states that if we perturb a system in the direction of a relevant operator, then we expect for the system to either lose conformality altogether or to flow to a different CFT. In the terminology of dynamical systems, standard RG trajectories are heteroclinic orbits. The classical example is the Wilson-Fischer fixed point: by carefully perturbing a Gaussian theory in $4-\epsilon$ dimension we flow to a weakly coupled interacting CFT, which in three dimensions interpolates to the Ising model. Homoclinic bifurcations provide exotic counterexamples to this general picture: if we perturb the system in the direction of a relevant operator, we come back to the original fixed point, which tentatively we can term a {\it homoclinic CFT}. Such RG behaviour obviously violates the $F$-theorem so that homoclinic fixed points must be non-unitary, as is generally the case for CFTs with symmetry groups of non-integer rank \cite{Binder:2019zqc}. 
	
	If we tune the bifurcation parameters so as to approach the BT point along the saddle homoclinic bifurcation (the red curve in figure \ref{fig:bifurcationDiagram}), then the homoclinic orbit shrinks to a point and vanishes. In this limit, the \textcolor{red}{red} homoclinic CFT and the \textcolor{green}{green} spooky fixed point merge and become a logarithmic CFT.
	
	\section{Zero-Hopf Bifurcations: The Road to Chaos}
	
	The Bogdanov-Takens bifurcation is not the only codimension-two bifurcation that can be observed to take place in the model \eqref{action}. The theory also possesses two points in the space of $g_i$, $M$, and $N$ where the stability matrix has a pair of purely imaginary eigenvalues and one zero eigenvalue. Such fixed points indicate what is known as a Zero-Hopf (ZH) or a Fold-Hopf bifurcation. This type of bifurcation was classified in \cite{takens1974singularities} and can be divided into six sub-types. In the notation of \cite{guckenheimer2013nonlinear}, the model has a type I ZH bifurcation at $(M,N)\approx (0.8447,-1.807)$ and a type IIa ZH bifurcation at $(M,N)\approx (-3.816,1.188)$. At a type I bifurcation point, a saddle-node bifurcation is incident to a pitchfork bifurcation, and there are no nearby cyclic orbits. At a type IIa point, a saddle-node bifurcation is again incident to a pitchfork bifurcation, but additionally a Hopf bifurcation is also incident to the point, except that the stability coefficient of the associated limit cycle (what was referred to as the Hopf constant in \cite{Jepsen:2020czw}) exactly vanishes in a quadratic approximation, so that cubic fluctuations or higher decide the fate of the cyclic flow near a type IIa point.
	
	Generally, ZH bifurcation points are of particular interest because it is known that in their vicinity what is known as a Shilnikov homoclinic orbit may develop and render the system chaotic \cite{kuznetsov2013elements,shilnikov1965case}. Recently it was proven in \cite{baldoma2020hopf} that the presence of ZH bifurcations of type III guarantees the existence of a Shilnikov orbit and a nearby infinite set of saddle periodic orbits. This nontrivial invariant set can be embedded in an attracting domain, thus implying Shilnikov chaos. 
	
	The ZH points of the model in the present paper are not of type III, and we cannot claim that the system is chaotic. It may be worthwhile to investigate if there exist other models that meet the simple criteria for the assured appearance of chaos.
	
	\section{Conclusion and Outlook}
	The approach suggested and adopted in \cite{Gukov:2016tnp,Kuipers:2018lux,Jepsen:2020czw} of studying the beta functions and renormalization of QFTs from the general perspective of dynamical systems provides a method of understanding the full range of possible RG flows. A powerful tool to this end is offered by Bogdanov's and Taken's bifurcation theorem \cite{hopf1942bifurcation}, which lists a simple set of conditions that guarantee the existence of a homoclinc RG orbit, and which can be checked already at first order in perturbation theory. 
	
	In this short letter we have presented a QFT that satisfies these conditions, namely a supersymmetric model with global symmetry group $O(N) \times O(M)$, where $N$ and $M$ play the role of the bifurcation parameters of the system. We determined a number of parameter values where a BT bifurcation takes place and investigated the nearby RG flow to uncover the presence of homoclinic orbits, where the perturbation of a fixed point by a relevant operator induces an RG flow that returns to its starting point along an irrelevant direction.
	
	There are several bifurcation theorems that give simple criteria for other novel kinds of RG flows \cite{guckenheimer2013nonlinear,kuznetsov2013elements,arnold2012geometrical}. Some of these theorems allow for the determination of the onset of chaotic flow based on straightforward computations around fixed points \cite{baldoma2020hopf}. It would be interesting to find out if QFTs give birth to chaos when $N$ becomes fractional.
	
	\section*{Acknowledgments}
	We are grateful to Igor R. Klebanov for very insightful discussions and suggestions throughout the project. We are also grateful to Alexander Gorsky, Alexei Milekhin, Yuri Kuznetsov, Alexander Polyakov, Sergey Gukov, Slava Rychkov, and Bernardo Zan for valuable discussions and comments. We also thank Maikel Bosschaert for spotting a considerable typo in appendix B.
	This research was supported in part by the US NSF under Grant No. PHY-1914860.

	\appendix

	\section{Transformation to Normal Form}
	\label{sec:deriv}
	In this appendix we present a case study of one of the Bogdanov-Takens bifurcations present in the QFT with action \eqref{action}.
	
	At $M=M^\ast \approx 0.2945$ and $N=N^\ast \approx 4.036$, there exists an RG fixed point $g^\ast$ such that stability matrix 
	\begin{align}
		M_i{}^j=\left.\frac{\partial \beta_i}{\partial g_j}\right|_{g^\ast} \label{eq:stabilityMatrix}
	\end{align}
	has a zero eigenvalue of multiplicity two in addition to two non-zero eigenvalues: $\lambda_1=2$ and $\lambda_2 \approx -2.357$ (working in units of $\epsilon$). This implies the existence of a matrix $V=\big(\vec{v}_1,\vec{v}_2,\vec{v}_3,\vec{v}_4\big)^\intercal$ such that
	\begin{gather}
		V^{-1}MV=
		\left(\begin{matrix}
			\lambda_1 & 0 & 0 & 0 \\
			0 & \lambda_2 & 0 & 0 \\
			0 & 0 & 0 & \lambda_3 \\
			0 & 0 & 0 & 0
		\end{matrix}\right)  \label{eq:numstabmatrix}
	\end{gather}
	where $\vec{v}_i$ are (generalized) unit eigenvectors, and $\lambda_3 \approx 7.555$ is a generalized eigenvalue. That is,
	\begin{gather*}
		M\vec{v}_{1,2}=\lambda_{1,2} \vec{v}_{1,2}\,, \quad
		%       M\vec{v}_2=\lambda_2 \vec{v}_2, \quad
		M\vec{v}_3=0\,, \quad
		M\vec{v}_4=\lambda_3 \vec{v}_3\,.
	\end{gather*}
	By a change of variables from $\vec{g}$ to $\vec{h}=V^{-1}(\vec{g}-\vec{g}^\ast)$, we obtain differential equations where $h_1$ and $h_2$ do not mix linearly with $h_3$ and $h_4$:
	\begin{gather}
		\beta_{h_{1,2}}=\lambda_{1,2}h_{1,2}+\mathcal{O}(h^2)\,.
	\end{gather}
	
	Consider now the case when $M=M^\ast+\delta M$ and $N=N^\ast+\delta M$, where $\delta M$ and $\delta N$ are each suppressed by a small parameter $\alpha \ll 1$. That is, $\delta M, \delta N\sim \alpha$. If we adopt the $h$ variables, $h_1$ and $h_2$ will now mix linearly with each other and with $h_3$ and $h_4$, and their beta functions will contain constant terms. We can write
	\begin{gather*}
		\frac{dh_1}{dt}=B_{0,1}+A_{1,1}h_1+B_{1,2}h_2+B_{1,3}h_3+B_{1,4}h_4+\mathcal{O}(h^2)\\
		\frac{dh_2}{dt}=B_{0,0}+B_{2,1}h_1+A_{2,2}h_2+B_{2,3}h_3+B_{2,4}h_4+\mathcal{O}(h^2)\,,
	\end{gather*}
	for some coefficients $A_{i,j}$ and $B_{i,j}$, where the coefficients $B_{i,j}$ are all suppressed in $\alpha$. Introducing new variables
	\begin{gather}
		y_1 = h_1+ C_{1,0} + C_{1,3}h_3 + C_{1,4}h_4\,, \nonumber \\
		y_2 = h_2+ C_{2,0} + C_{2,3}h_3 + C_{2,4}h_4 \,,
	\end{gather}
	we can choose the six coefficients $C_{i,j}$ such that $\frac{dy_{1,2}}{dt}$ do not contain constant terms nor mix linearly with $h_3$ and $h_4$. This means that, studying only the RG flow near the origin so that cubic terms and higher can be disregarded, the surface $y_1=y_2=0$ is an invariant manifold. On this surface, we can define variables $y_3 = h_3$ and $y_4 = \lambda_3 h_4$, whose RG flow to quadratic order in the dynamic variables is governed by the differential equations
	\begin{gather}
		\frac{dy_3}{dt}=a_{00}+a_{10}y_3+a_{01}y_4
		+\frac{1}{2}a_{20}y_3^2
		+a_{11}y_3y_4+\frac{1}{2}a_{02}y_4^2\,, \label{A7}
		\\
		\frac{dy_4}{dt}=b_{00}+b_{10}y_3+b_{01}y_4
		+\frac{1}{2}b_{20}y_3^2+b_{11}y_3y_4+\frac{1}{2}b_{02}y_4^2\,. \nonumber
	\end{gather}
	For the specific BT point we are considering, omitting higher order terms in $\alpha$, the values of the various coefficients are given by
	\begin{gather}
		a_{00}=\, 12.75\, \delta N + 2.618\, \delta M\,,
		\notag
		\\
		a_{10}=\, 46.93\, \delta N -8.164\, \delta M\,, \notag\\
		a_{01}=\, 1\,,\quad
		a_{20}=\, 0.6040\,, \notag\\
		a_{11}=\, -2.268\,, \quad
		a_{02}=\, -0.9708\,, \notag 
		\\
		b_{00}=\, 19.49\, \delta N -6.753\, \delta M
		-638.9\, \delta N^2 \notag\\
		+99.89\, \delta M^2-71.61\, \delta N\,\delta M\,, \notag \\
		b_{10}=\,  -51.76\, \delta N +17.38\, \delta M\,, \notag \\
		b_{01}=\,  -18.05\, \delta N -0.4165\, \delta M\,, \notag \\
		b_{20}=\, 2.775\,,\quad
		b_{11}=\, -0.7935\,, \quad 
		b_{02}=\, -1.868 \,.
	\end{gather}
	We now restrict attention to the case when $\delta N$ and $\delta M$ are tuned such that $b_{00}\sim \alpha^2$, for the reason that this will turn out to be the regime where the existence of a homoclinic orbit can be reliably established. Working merely to leading order in $\alpha$ and to quadratic order in the dynamical variables, we perform a reparametrization from $t$ to $\tau$ via 
	\begin{align}
		d\tau=& -\frac{1}{2}\frac{b_{20}}{a_{20}+b_{11}}\Big(1-\frac{2a_{11}+b_{02}}{2}\,\frac{dy_3}{dt}\Big)^{-1}dt
	\end{align}
	and change to variables $\eta_1$ and $\eta_2$ given by
	\begin{gather}
		\eta_1 = 2\frac{a_{20}+b_{11}}{b_{20}}\Big((a_{20}+b_{11})y_3+a_{01}+b_{01}-a_{00}a_{11}-a_{00}b_{02}\Big)   \nonumber
		\\
		\eta_2= -4\frac{(a_{20}+b_{11})^3}{b_{20}^2}\times
		\\
		\times
		\bigg(
		1-\frac{2a_{11}+b_{02}}{2}\Big(y_3+\frac{a_{10}b_{01}-a_{00}a_{11}-a_{00}b_{02}}{a_{20}+b_{11}}\Big)
		\bigg)\frac{dy_3}{dt}\,, \nonumber
	\end{gather}
	where for $\frac{dy_3}{dt}$ one should substitute the RHS of \eqref{A7}. In these new variables, the differential equations of the dynamical system are brought into the normal form introduced by Bogdanov: 
	\begin{gather}
		\left\{\begin{matrix*}[l]
			\dot{\eta}_3=\eta_4\,, \\
			\dot{\eta}_4 = \delta_1+\delta_2 \eta_3+\eta_3^2 + \eta_3 \eta_4\,,\\
		\end{matrix*}\right. \label{eq:A10}
	\end{gather}
	where we have omitted terms of cubic order and higher in $\eta$, and the parameters $\delta_1$ and $\delta_2$ are given by
	\begin{widetext}
		\begin{gather}
			\delta_1 =\frac{8(a_{20}+b_{11})^4}{b_{20}^3}
			\bigg(
			b_{00}-a_{00}b_{01}+\frac{a_{00}^2b_{02}}{2}
			+\frac{\big(a_{10}+b_{01}-a_{00}(a_{11}+b_{02})\big)(a_{00}b_{11}-b_{10})}{a_{20}+b_{11}}
			+\frac{\big(a_{10}+b_{01}-a_{00}(a_{11}+b_{02})\big)^2b_{20}}{2(a_{20}+b_{11})^2}
			\bigg)\,,
			\notag\\
			\delta_2  
			=\frac{4(a_{20}+b_{11})}{b_{20}^2}\bigg((a_{20}+b_{11})(b_{10}-a_{00}b_{11})
			-\big(a_{10}+b_{01}-a_{00}(a_{11}+b_{02})\big)b_{20}\bigg)\,.
		\end{gather}
	\end{widetext}

	The transformations by which we arrived at the equations \eqref{eq:A10} can be applied more generally to dynamical systems where the stability matrix contains a Jordan block with zero eigenvalues, as long as the following conditions are met: 
	
	\begin{gather}
		\lambda_3\neq 0\,, \quad  b_{20}\neq 0\,,\quad a_{20}+b_{11} \neq 0\,. \label{eq:BTcond}    
	\end{gather}
	
	This fact is known as the Bogdanov-Takens bifurcation theorem \cite{bogdanov,takens2001forced,MR2197746,MR3501359}. The precise formulation of the theorem is this:
	
	\begin{theorem}
		Suppose we have a system of differential equations
		\begin{equation}
			\dot{\vec{x}} = \vec{f}(\vec{x},\vec{\alpha}), \quad \vec{x} \in \mathbb{R}^n, \, \vec{\alpha} \in \mathbb{R}^2, 
		\end{equation}
		where $\vec{f}$ is smooth, and suppose further that $\vec{f}(0,0)=0$ and that the stabililty matrix $\left(\frac{\partial \vec{f}}{\partial \vec x}\right)\Big|_{\vec{x}=\vec{\alpha}=0}$ has a Jordan cell of size two with zero eigenvalues: %algebraically double zero eigenvalue:
		$\lambda_{1,2}=0$ and other eigenvalues  $\lambda_3,\ldots,\lambda_n$ with non-zero real parts. Assume that the map $\left(\vec{x},\vec{\alpha}\right) \longmapsto \left(\vec{f}(\vec{x},\vec{\alpha}), \tr\left(\frac{\partial f_i}{\partial x_j}\right) ,  \det\left(\frac{\partial f_i}{\partial x_j}\right) \right)$ is smooth and that the non-degeneracy conditions \eqref{eq:BTcond} are satisfied. Then there exists a smooth invertible variable transformation, a direction-preserving time reparametrization,
		and a smooth invertible change of parameters that together reduce the system
		to the normal form \eqref{eq:BTsys},
		where $\delta_{1,2}$ are functions of $\vec{\alpha}$ and $s=\pm 1$.
	\end{theorem}
	
	Furthermore, there is a theorem stating that the suppressed terms of cubic order and higher in \eqref{eq:BTsys} do not change the local topology of the flow. But the topology of the flow of the normal form system, omitting cubic and higher terms, is well understood. In particular, it is known that depending on the values of $\delta_1$ and $\delta_2$, the flow near the origin can be divided into four distinct regions. As described in section \ref{sec:BT}, these four regions are separated by codimension-one bifurcations located on the curve $\delta_1=\frac{1}{4}\delta_2^2$ and, for $\delta_2<0$, on the curves $\delta_1=0$ and $\delta_1=-\frac{6}{25}\delta_2^2$. For the specific BT point at $M\approx 2.945$ and $N\approx 4.036$, we can translate these equations into relations between $\delta M$ and $\delta N$, whereby we arrive at the picture of figure \ref{fig:bifurcationDiagram}.
	
	As mentioned in section \ref{sec:model}, there is a fixed point at $M=2$, $N=3$, which does have a stability matrix with a zero eigenvalue of multiplicity two, but which does not meet the genericity conditions \eqref{eq:BTcond}. Because the point is situated at the value $N=3$, the metric is degenerate, and only three of the four operators \eqref{eq:Ops} are linearly dependent. Specifically,
	\begin{align}
		2O_1+4O_2-2O_3-O_1\,\big|_{N=3}=0\,. 
	\end{align}
	It turns out that the operator $\mathcal{O}_2$ corresponding to the eigenvalue $\lambda_2$ is identically zero owing to this linear relationship, so that the space of couplings is three-dimensional, and the stability matrix at the fixed point has eigenvalues $2$, $0$, and $0$. But the fixed point is not described by a logarithmic CFT, nor does it represent a BT bifurcation, the reason being that the generalized eigenvalue $\lambda_3$ is zero.
	
	\section{Two-Loop Beta Functions}
	\label{sec:betas}
	The beta functions of the four coupling constants admit loop expansions
	\begin{align}
		\beta_i = -\epsilon g_i+\beta_i^{(2)}+\mathcal{O}(g^5)\,, \label{eq:b1}
	\end{align}
	where $\beta_i^{(2)}$ denotes the two-loop contributions, which are cubic in the couplings. By explicit computation using \cite{Popov:2019nja}, we find that these are given by

	\begin{widetext}
		\begin{align}\label{b1}
			\beta_1^{(2)}=\,&\frac{1}{8\pi^2 N^2}\bigg(
			32 g_1^2 g_4 N (-40 - 8 M + 8 N + 7 M N + 8 N^2) + 
			16 g_1^2 g_3 N (-80 - 24 M + 30 N + 4 M N + 7 N^2 + 4 M N^2)  
			\nonumber  \\ &
			+ 
			16 g_1 g_3^2 N^2 (32 + 2 M + N + M N + N^2 + M N^2) + 
			64 g_1 g_2 g_4 N (-32 - 4 M + 16 N + 2 M N + 5 N^2 + 2 M N^2) 
			\nonumber  \\ &
			+ 
			4 g_2^2 g_3 N (-256 - 64 M + 72 N + 10 M N + 19 N^2 + 2 M N^2)
			+ 
			64 g_1 g_4^2 N^2 (22 - 2 M + M N + M N^2) 
			\nonumber  \\ &
			+ 
			16 g_1 g_2 g_3 N (-144 - 40 M + 42 N + 17 M N + 21 N^2 + 5 M N^2) + 
			128 g_1 g_3 g_4 N^2 (3 + 5 M + N + N^2)
			\nonumber  \\ &
			+ 
			g_2^3 (896 + 128 M - 352 N - 48 M N - 12 N^2 - 12 M N^2 + 7 N^3 + 
			2 M N^3) +96g_2^2 g_4 (-8 + N) N
			\\ &
			+ 
			4 g_1^2 g_2 (928 + 224 M - 392 N - 100 M N - 2 N^2 + 11 M N^2 + 
			23 N^3 + 7 M N^3 + 4 N^4) + 768 g_2 g_3 g_4 N^2
			\nonumber \\ &
			+ 
			4 g_1^3 (352 + 96 M - 152 N - 44 M N + 10 N^3 + 3 M N^3 + M N^4) + 
			16 g_2 g_3^2 N^2 (16 + 7 M + N + N^2) 
			\nonumber  \\ &
			+ 
			2 g_1 g_2^2 (1600 + 320 M - 656 N - 136 M N + 32 N^2 + 10 M N^2 + 
			50 N^3 + 10 M N^3 + 5 N^4 + 2 M N^4)
			\bigg)\,,
			\nonumber
		\end{align}
		
		\begin{align}\label{b2}
			\beta_2^{(2)}=\,&\frac{1}{8\pi^2 N^2}\bigg(
			64 g_1 g_2 g_4 N (-80 - 16 M + 16 N + 5 M N + 7 N^2) 
			+ 
			16 g_2 g_3^2 N^2 (48 + 9 M + 2 N + M N + 2 N^2 + M N^2) 
			\nonumber \\ &
			+ 
			16 g_1^2 g_3 N (-128 - 32 M + 32 N + 12 M N + 14 N^2 + 3 M N^2)
			+ 128 g_2 g_3 g_4 N^2 (9 + 5 M + N + N^2)
			\nonumber  \\ &
			+ 
			16 g_1 g_2 g_3 N (-272 - 72 M + 82 N + 15 M N + 24 N^2 + 6 M N^2) 
			+ 192 g_1^2 g_4 N (-12 - 2 M + 4 N + N^2) 
			\\ &
			+  16 g_2^2 g_4 N (-176 - 40 M + 58 N + 14 M N + 17 N^2 + 11 M N^2) 
			+  32 g_1 g_3^2 N^2 (16 + 7 M + N + N^2) 
			\nonumber  \\ &
			+  4 g_2^2 g_3 N (-576 - 160 M + 176 N + 54 M N + 74 N^2 + 17 M N^2) 
			+  64 g_2 g_4^2 N^2 (22 - 2 M + M N + M N^2)
			\nonumber  \\ &
			+ 
			8 g_1^3 (288 + 64 M - 112 N - 24 M N - 6 N^2 + 3 M N^2 + 5 N^3 
			+   2 M N^3 + N^4) 
			+1536 g_1 g_3 g_4 N^2 
			\nonumber     \\ &
			+  2 g_1 g_2^2 (3968 + 1024 M - 1600 N - 416 M N - 56 N^2 - 22 M N^2 + 
			85 N^3 + 17 M N^3 + 11 N^4) 
			\nonumber    \\ &
			+  4 g_1^2 g_2 (1856 + 448 M - 736 N - 176 M N - 22 N^2 - 5 M N^2 + 
			43 N^3 + 8 M N^3 + 3 N^4 + 2 M N^4) 
			\nonumber     \\ &
			+  g_2^3 (2816 + 768 M - 1152 N - 320 M N + 12 N^2 - 12 M N^2 + 69 N^3 + 
			30 M N^3 + 7 N^4 + 5 M N^4)
			\bigg)\,, \nonumber
		\end{align}
		
		\begin{align}
			\beta_3^{(2)}=\,&\frac{1}{8\pi^2 N^3}\bigg(
			32 g_3^2 g_4 N^3 (18 + 14 M + 7 N + 7 N^2)  + 
			96 g_1^2 g_4 N (8 + 2 N^2 + M N^2) +384 g_1 g_2 g_4 N (4 + N^2)
			\nonumber \\ &
			+ 
			24 g_3^3 N^3 (16 + 2 M + 2 N + M N + 2 N^2 + M N^2) + 
			64 g_2 g_3 g_4 N^2 (-20 - 4 M + 10 N + 2 M N + 5 N^2 + 2 M N^2) 
			\nonumber \\ &
			+ 
			16 g_1 g_3^2 N^2 (-52 - 14 M + 26 N + 7 M N + 14 N^2 + 7 M N^2) 
			+ 
			128 g_1 g_3 g_4 N^2 (-10 - 2 M + 5 N + M N + 5 N^2) 
			\nonumber \\ &
			+ 
			8 g_2 g_3^2 N^2 (-104 - 28 M + 52 N + 14 M N + 38 N^2 + 7 M N^2) + 
			64 g_3 g_4^2 N^3 (22 - 2 M + M N + M N^2)
			\nonumber \\ &
			+ 
			16 g_1 g_2 g_3 N (208 + 56 M - 48 N - 12 M N + 12 N^2 + 3 M N^2 + 
			8 N^3 + M N^3 + 2 N^4)  + 96 g_2^2 g4 N (8 + N^2) 
			\nonumber    \\ &
			+ 
			8 g_1^2 g_3 N (208 + 56 M - 48 N - 12 M N + 12 N^2 + 6 M N^2 + 4 N^3 + 
			3 M N^3 + M N^4) 
			\\ &    
			+ 
			8 g_1^3 (-96 - 16 M + 24 N + 4 M N - 20 N^2 - 10 M N^2 + 6 N^3 + 
			2 M N^3 + N^4 + M N^4) 
			\nonumber    \\ &
			+ 
			4 g_2^2 g_3 N (416 + 112 M - 96 N - 24 M N + 18 N^2 + 6 M N^2 + 
			12 N^3 + 4 M N^3 + 2 N^4 + M N^4) 
			\nonumber   \\ &
			+ 
			g_2^3 (-768 - 128 M + 192 N + 32 M N - 96 N^2 + 32 N^3 + 4 M N^3 + 
			7 N^4 + 2 M N^4) 
			\nonumber   \\ &
			+ 
			2 g_1 g_2^2 (-1152 - 192 M + 288 N + 48 M N - 176 N^2 - 40 M N^2 + 
			70 N^3 + 10 M N^3 + 21 N^4 + 2 M N^4) 
			\nonumber   \\ &
			+ 
			4 g_1^2 g_2 (-576 - 96 M + 144 N + 24 M N - 104 N^2 - 40 M N^2 + 
			34 N^3 + 13 M N^3 + 11 N^4 + 3 M N^4)
			\bigg)\,,
			\nonumber
		\end{align}
		
		\begin{align}\label{b4}
			\beta_4^{(2)}=\,&\frac{1}{8\pi^2 N^3}\bigg(
			8 g_1^2 g_3 N (80 + 16 M - 12 N + 4 N^2 + 2 M N^2 + 3 N^3 + N^4)
			\nonumber   \\ &
			+ 224 g_1 g_4^2 N^2 (-4 - 2 M + 2 N + M N + 2 N^2) 
			+  16 g_1 g_3^2 N^2 (-16 - 2 M + 8 N + M N + 7 N^2)
			\nonumber   \\ &
			+  96 g_4^3 N^3 (8 - 2 M + M N + M N^2) 
			+  64 g_2 g_3 g_4 N^2 (-14 - 4 M + 7 N + 2 M N + 6 N^2 + M N^2) 
			\nonumber   \\ &
			+  224 g_2 g_4^2 N^2 (-4 - 2 M + 2 N + M N + N^2 + M N^2) 
			+  32 g_3^2 g_4 N^3 (22 + N + M N + N^2 + M N^2) 
			\nonumber   \\ &
			+  64 g_1 g_3 g_4 N^2 (-14 - 4 M + 7 N + 2 M N + 2 N^2 + 2 M N^2)
			\nonumber   \\ &
			+  8 g_2 g_3^2 N^2 (-32 - 4 M + 16 N + 2 M N + 9 N^2 + 2 M N^2)
			+ 24 g_3^3 N^3 (4 + 2 M + N + N^2) 
			\nonumber   \\ &
			+  16 g_1 g_2 g_3 N (80 + 16 M - 12 N + 7 N^2 + 2 M N^2 + N^3)
			+ 224 g_3 g_4^2 N^3 (2 M + N + N^2) 
			\nonumber   \\ &
			+  4 g_2^2 g_3 N (160 + 32 M - 24 N + 17 N^2 + 7 M N^2 + 4 N^3 + N^4) 
			\nonumber   \\ &
			+  32 g_1 g_2 g_4 N (96 + 36 M - 24 N - 12 M N + 2 N^2 - 2 M N^2 + 4 N^3 + 
			M N^3 + N^4) 
			\nonumber   \\ &
			+  4 g_1^3 (-160 - 48 M + 40 N + 12 M N + 4 N^2 + 2 M N^2 + 4 N^3 + 
			M N^3 + 2 N^4) 
			\nonumber   \\ &
			+  2 g_1 g_2^2 (-960 - 288 M + 240 N + 72 M N - 88 N^2 - 20 M N^2 + 
			39 N^3 + 7 M N^3 + 13 N^4)
			\nonumber   \\ &
			+  16 g_1^2 g_4 N (96 + 36 M - 24 N - 12 M N - 4 N^2 - 2 M N^2 + 2 N^3 + 
			3 M N^3 + M N^4) 
			\nonumber   \\ &
			+  8 g_1^2 g_2 (-240 - 72 M + 60 N + 18 M N - 8 N^2 - M N^2 + 6 N^3 + 
			2 M N^3 + N^4 + M N^4) 
			\nonumber   \\ &
			+  8 g_2^2 g_4 N (192 + 72 M - 48 N - 24 M N + 10 N^2 + 2 M N^2 + 6 N^3 + 
			4 M N^3 + N^4 + M N^4) 
			\nonumber   \\ &
			+  g_2^3 (-640 - 192 M + 160 N + 48 M N - 96 N^2 - 24 M N^2 + 36 N^3 + 
			12 M N^3 + 10 N^4 + 5 M N^4)
			\bigg)\,.
			\nonumber
		\end{align}

	\end{widetext}
	By proper rescaling $g_i \to \eps \tilde{g_i}$ we can get rid of $\eps$ in the equation \eqref{eq:b1}. It can be checked that there exists a function $F$ of the couplings such that the beta-functions \eqref{b1}-\eqref{b4} can be cast in the form
	\begin{gather}
		\beta_i = G_{ij} \frac{\partial F}{\partial g_j}
	\end{gather}
	where the metric has the components
	\begin{widetext}
		\begin{align*}
			G_{11}=&\frac{M N^4+3M N^3+2N^3-2MN^2-4N^2-12MN-24N+24M +48}{2N^2}\,,
			\\
			G_{12}=&\frac{N^4+MN^3+4N^3-2MN^2-4N^2-12MN-24N+24M+48}{2N^2}\,,
			\\
			G_{13}=&2\,\frac{MN^2+N^2+MN+2N-2M-4}{N}\,,\quad G_{14}=2\,\frac{2N^2+MN+2N-2M-4}{N}\,,
			\\
			G_{22}=&\frac{MN^4+N^4+4MN^3+6N^3-4MN^2-8N^2-24MN-48N+48M+96}{4N^2}\,,
			\\
			G_{23}=&\frac{MN^2+3N^2+2MN+4N-4M-8}{N}\,,\quad
			G_{24}=2\frac{MN^2+N^2+MN+2N-2M-4}{N}\,,
			\\
			G_{33}=&\frac{MN^3+N^3+MN^2+N^2+4N}{N}\,,\quad
			G_{34}=2(N^2+N+2M)\,, \quad 
			G_{44}=2(4-2M+MN+MN^2)\,.
		\end{align*}
	\end{widetext}
	Taking the determinant of this metric, one arrives at \eqref{eq:det}.
	
	\section{Bifurcation Conditions}
	
	When working to two-loop level, the beta functions have the property that $\beta_i + \epsilon g_i $ is a homogeneous function of degree three. Hence, by Euler's theorem,
	\begin{align}
		g_j\frac{\partial (\beta_i+\epsilon g_i)}{\partial g_j} = 3(\beta_i+\epsilon g_i)\,.
	\end{align}
	By evaluating this equation on an RG fixed point $\vec{g}^*$, one finds that
	\begin{gather}
		\left.M_j{}^i\right|_{\vec{g}=\vec{g}^*}g^*_i=2 \eps g_j^*\,,
	\end{gather}
	where $M_j{}^i = \frac{\partial \beta_j}{\partial g_i}$ is the stability matrix defined in \eqref{eq:stabilityMatrix}. Hence, for any non-trivial fixed point, the stability matrix has an eigenvalue $\lambda_1=2\eps$ equal to two. This fact allows for simplifications in the conditions for bifurcations to occur.
	
	Consider the determinant $\det\left(M-\lambda \right)$. Since it is a fourth order polynomial in $\lambda$, we can write
	\begin{align}
		\det\left(M-\lambda\right) = \lambda^4+A\lambda^3+B\lambda^2+C\lambda+D\,,
	\end{align}
	where $A$, $B$, $C$, and $D$ are $M$- and $N$-dependent polynomials in the coupling constants. Specifially
	\begin{align}
		A = - \text{Tr}M\,,\quad B=M_{ij,ij}\,,\quad C=-M_{i,i}\,,\quad D=\det M\,,
		\nonumber
	\end{align}
	where $M_{i,j}$ and $M_{ij,kl}$ are the first and second minors of the stability matrix. But when evaluated on a fixed point, we also have the following factorization in terms of eigenvalues (in units where $\epsilon=1$)
	\begin{align}
		\det\left(M_\ast-\lambda \right) = (\lambda -2)(\lambda -\lambda_2)(\lambda -\lambda_3)(\lambda -\lambda_4)\,. \nonumber
	\end{align}
	By expanding out the factors on the RHS, we can relate $A$, $B$, and $C$ to the eigenvalues.
	
	A {\bf saddle-node bifurcation} occurs when at a fixed point we have a zero-eigenvalue: $\lambda_2 =0$. In this case, one finds that
	\begin{align}
		&A=-2-\lambda_3-\lambda_4\,, \hspace{5mm}
		B=2(\lambda_3+\lambda_4)+\lambda_3\lambda_4\,, \nonumber
		\\
		&C=-2\lambda_3\lambda_4\,, 
	\end{align}
	from which one can derive the equation
	\begin{align}
		2B+C=-4(A+2)\,. \label{eq:saddleNodeCondition}
	\end{align}
	Of course we also have the equation $D=0$, but the condition \eqref{eq:saddleNodeCondition} is easier to check numerically on account of the many terms in the determinant.
	
	At a {\bf Hopf bifurcation}, we have a conjugate pair of imaginary eigenvalues:
	\begin{align}
		\lambda_3=i\chi\,, \hspace{6mm} \lambda_4=-i\chi\,,
	\end{align}
	where $\chi$ is a real number. Consequently we find that
	\begin{align}
		A=-2-\lambda_2\,, \hspace{3.7mm}
		B-2\lambda_2+\chi^2\,, \hspace{3.7mm}
		C=-(2+\lambda_2)\chi^2\,. \nonumber
	\end{align}
	From these equations, we derive the condition
	\begin{align}
		C=A(B+4+2A)\,.  
	\end{align}

	At a {\bf Bogdanov-Takens bifurcation}, there are two zero eigenvalues: $\lambda_3=\lambda_4=0$. From this, one obtains the conditions
	\begin{align}
		B=-2(A+2)\,, \hspace{8mm} C=0\,.
	\end{align}
	
	We have a {\bf Zero-Hopf bifurcation} when
	\begin{align}
		\lambda_2=0\,, \hspace{5mm}
		\lambda_3=i\chi\,, \hspace{5mm}
		\lambda_4=-i\chi\,,
	\end{align}
	with $\chi$ a real number. In this case
	\begin{align}
		A= -2 \,, \hspace{8mm} C=-2B\,.
	\end{align}
	
	\bibliography{biblio}
\end{document}